\documentclass{elsart}

\usepackage[english]{babel}
\usepackage{dcolumn}
\usepackage{amsmath,amssymb,epsfig,float}

\begin{document}
\begin{frontmatter}
\title{Probing extra dimensions with higher dimensional black hole analogues?}
\author[apctp]{Xian-Hui Ge}~~~
\author[apctp,kun]{Sung-Won Kim}
\address[apctp]{Asia-Pacific Center for Theoretical Physics,\\
 Pohang 790-784, Korea }
\address[kun]{
Department of Science Education, Ewha Woman's University, Seoul
120-750, Korea}
\thanks[email]{e-mail:gexh@www.apctp.org}
\begin{abstract}
  \hspace*{7.5mm}We propose that extra dimensions might be detected with higher
dimensional analogues of black holes. The usual 4-dimensional
acoustic(sonic)black hole metric is extended to arbitrary
dimensions.  The absorption cross-section of Hawking radiation on
the brane and in the bulk are calculated in the semiclassical
approximation.
\\

\end{abstract}
\end{frontmatter}

\section{Introduction}
\hspace*{7.5mm}A lot of interest in recent years has been raised for
field theories where the standard model of high-energy physics is
assumed to live on a 3-brane embedded in a larger space-time, while
only the gravitational fields are in contrast usually considered to
live in the whole spacetime \cite{horava,Arkani,randall,dvali}.
Arkani-Hamed \textit{et al} \cite{Arkani}, proposed a new framework
for solving the hierarchy problem which does not rely on
supersymmetry or techi-color. The novelty in this idea was that the
traditional picture of Planck-length-sized additional spacelike
dimensions ($l_{p}\simeq 10^{-33}$cm) was abandoned, and the extra
dimensions could have a size as large as 1 mm. The fact that we do
not see experimental signs of the extra dimensions despite that the
compactification scale of the extra dimensions $\mu_{c}\sim
1/{V_{n}^{1/n}}$ would have to be much smaller than the weak scale,
implies that only gravity can propagate in the extra-dimensional
spacetime and all ordinary matter: electromagnetic, weak and strong
forces, is restricted to live on a (3+1) dimensional hypersurfaces,
a 3-brane.\\
  \hspace*{7.5mm}\hspace*{7.5mm} The presence of extra dimensions in brane world
gravity models will inevitably change the properties and physics of
black holes. In brane-world scenario, the production cross-section
for black holes is greatly enhanced. Detectable signals of black
holes might be found via Hawking evaporation to brane-localized
modes. If the scale of quantum gravity is near a Tev, the
  Large Hadron Collider (LHC) will become a ``black hole factory'' \cite{savas}.
  The correlations between the black hole
  mass and its temperature, deduced from the energy spectrum of the
  decay products, can test Hawking's evaporation law. The absorption cross-section
  (and also grey body factors) of black holes in different physical scenario have been
investigated by several authors\cite{page,unr,pkanti,kmr,kanti}. The
emission rate of 4-dimensional acoustic black holes for $S$ wave has
been studied in Ref.\cite{kim}. In this paper,
we would like to investigate the absorption cross-section of higher dimensional acoustic black holes.\\
\hspace*{7.5mm}The remarkable work of Unruh, in 1981, developed a
way of mapping certain aspects of black holes in supersonic flows
and pointed out that propagation of sound in a fluid or gas turning
supersonic~\cite{unruh}, is similar to the propagation of a scalar
field close to a black hole, and thus experimental investigation of
the Hawking radiation is possible. From then on, several candidates
have been considered for the experimental test of the analogue of
black holes~\cite{novello}.\\
\hspace*{7.5mm}In the present study, we extend 4-dimensional
acoustic black hole metric into arbitrary dimensional space-time in
Section 2. Then following the methods of Ref.\cite{kmr},
 we calculate the corresponding absorption probabilities and
 energy emission rates of such black hole analogues. In section 3, we present the calculation of
bulk emission in $(4+n)$-dimensions for $l\geq 0$ case  by using the
low energy perturbation method. Section 4 performs the calculation
of brane-localized scalar emission, where the brane is embedded in a
$(4+n)$-dimensional bulk. We give the conclusions in Section 5.
\section{Metric of higher dimensional black hole analogues}
\hspace*{7.5mm}Since the dynamics of effective black hole analogues
metric are not described by the laws of gravity (i.e. Einstein
equations) in general, to extend the four dimensional effective
metric we should start with the equations of fluid dynamics. We
consider one kind of fluids which is permitted to propagate in
$(4+n)$-dimensional space-time. The fluid might be condensate states
of gravitons or some  higher-dimensional scalar particles. In the
following of our discussions, we simply neglect quantum effects of
such special quantum fluids and assume they obey the classical
fluids equation (Euler equation) in the low energy approximation.
The fundamental equations of fluid dynamics in $(4+n)$-dimensional
flat spacetime are the equation of continuity
\begin{equation}\partial_{t}\rho+\nabla\cdot(\rho\vec{v})=0,\end{equation}
and Euler's equations
\begin{equation}\rho\frac{d\vec{v}}{dt}\equiv \rho[\partial_{t}
\vec{v}+(\vec{v}\cdot \nabla)\vec{v}]=\vec{F},\end{equation} where
$\nabla=\hat{e}_{x}\frac{\partial}{\partial
x}+\hat{e}_{y}\frac{\partial}{\partial
y}+\hat{e}_{z}\frac{\partial}{\partial
z}+...+\hat{e}_{n}\frac{\partial}{\partial n}$, and $\vec{F}=-\nabla
p-\rho\nabla \phi$. Here $\phi$ denotes the Newtonian gravitational
potential. It is also assumed that the fluid to be irrotational and
inviscid, which imply $\nabla\times\vec{v}=0$. These equations can
be linearized in the vicinity of some mean flow solution with
$\rho=\rho_{0}+\varepsilon \rho_{1}+O(\varepsilon^2)$,
$p=p_{0}+\varepsilon
p_{1}+O(\varepsilon^2)$,$\varphi=\varphi_{0}+\varepsilon
\varphi_{1}+O(\varepsilon^2)$, redefining the fields as
${\nabla}h=\frac{\nabla p}{\rho}$, and $\vec{v}=\nabla\varphi$. One
can finally obtain the wave equation \cite{visser},
\begin{eqnarray}-\partial_{t}\left[\frac{\partial \rho}{\partial p}\rho_{0}
(\partial_{t}\varphi_{1}+\vec{v}_{0}\cdot \nabla\varphi)\right]
+\nabla\cdot\left[\rho_{0}\nabla\varphi_{1}-\frac{\partial \rho}
{\partial p}\rho_{0}\vec{v}_{0}(\partial_{
t}\varphi_{1}+\vec{v}_{0}\cdot\nabla\varphi_{1})\right]=0\end{eqnarray}
The above equation is identified with a massless scalar field
equation describing the sound wave in the curved spacetime
background
\begin{equation}
\label{klein}
\frac{1}{\sqrt{-g}}\partial_{\mu}(\sqrt{-g}g^{\mu\nu}\partial_{\nu}\varphi_{1})=0\end{equation}
with the background metric,
$g_{\mu\nu}=\left(\frac{\rho_{0}}{c}\right)^{\frac{2}{n+2}}\left(\begin{array}{cc}
  -c^2 +v_{0}^{2} & -v_{0}^{i}  \\
  -v_{0}^{j}& \delta_{ij}
 \end{array}\right)$, which is a $(n+4)\times (n+4)$ matrix, where
 the local speed of sound is defined by $c^{2}\equiv \frac{\partial p}{\partial
 \rho}$, and $i,j=1...n$.
 In spherical coordinates, assuming $v_{r}\neq0$,
 $v_{\theta_{1}}=v_{\theta_{2}}=...=0$, we then have
 \begin{eqnarray}
&&ds^{2}=
\left(\frac{\rho_{0}}{c}\right)^{\frac{2}{n+2}}\left[-c^{2}(1-\frac{v_{r}^2}{c^2})d\tau^{2}+(1-\frac{v_{r}^2}{c^2})^{-1}dr^{2}
+r^{2}d\Omega^{2}_{n+2}\right],
 \end{eqnarray} which is similar to the $n$-dimensional Schwarzschild
 black hole metric~\cite{myers},
 \begin{eqnarray}&&ds^{2}=\left[-\textsc{c}^{2}\left(1-\frac{r_{H}^{n+1}}{\textsc{c}^{2}r^{n+1}}\right)dt^2
  +\left(1-\frac{r_{H}^{n+1}}{\textsc{c}^{2}r^{n+1}}\right)^{-1}dr^2+
 r^2 d\Omega^{2}_{n+2}\right],\end{eqnarray}
 where $\textsc{c}$ is the light velocity and $r_{H}$ is the event horizon radius.
 The properties of higher dimensional Schwarzschild
 black holes in brane-world scenario have been discussed in
 Ref.\cite{arg}. However, the properties of acoustic black holes do
 not necessarily relate to the outside geometry of space-time since the dynamics of
 acoustic black holes do not obey Einstein equation.
  In Ref.\cite{ge}, we have discussed black hole analogues
 in brane-world scenario under some assumptions and found the
 properties of acoustic black holes in brane-world scenario are similar to real black holes.
  If $\rho$ is time and position independent, the  continuity equation ${\nabla}\cdot \vec{v}=0$ then
 implies that $v_{r}\propto \frac{1}{r^{n+2}}$. Because of the barotropic assumption, $\rho$ is position independent
 implying the pressure $p$ and the speed of sound $c$ are also position independent. We can define a normalization
 constant $r_{0}$ and set $v_{r}= \mathcal{C}/r^{n+2}$ with $\mathcal{C}=c r_{0}^{n+2}$.  $r_{0}$ is a parameter which can be determined
 by experiments. The metric can be
 rewritten as
 \begin{equation}
\label{acoustic}
d\tilde{s}^{2}=\left[-c^{2}\left(1-\frac{{r_{0}}^{2n+4}}{{r}^{2n+4}}\right)d\tau^{2}
+\left(1-\frac{{r_{0}}^{2n+4}}{{r}^{2n+4}}\right)^{-1}dr^{2}
+r^{2}d\Omega^{2}_{n+2}\right],
 \end{equation}where
 $d\tilde{s}^{2}=(\frac{c}{\rho_{0}})^{\frac{2}{n+2}}ds^2$.
 The temperature of an acoustic(sonic) black hole
 is given by $T=\frac{\hbar}{2\pi k}\left|\frac{\partial v_{r}}{\partial
 r}\right|_{v_{r}=c}$, where $v_{r}=c$ marks the exact position of the event horizon. For 4-dimensional cases,
 $v_{r}=c\frac{{r'}_{0}^{2}}{r^2}$, we have the numerical
 expression,
 \begin{equation}
T=6\times 10^{-7} K [c/300 \textrm{m/sec}][1\textrm{ mm}/r'_{0}].
 \end{equation}
 For (4+n)-dimensional cases,
 $v_{r}=c\frac{{r}_{0}^{n+2}}{r^{n+2}}$, the temperature can then be
 written as
 \begin{equation}
T=3(n+2)\times 10^{-7} K [c/300 \textrm{m/sec}][1
\textrm{mm}/r_{0}].
 \end{equation}
 \hspace*{7.5mm}It is clear from the above equations that unless $c$ is very large (i.e. to be the velocity of light $c=3\times10^{8}$
 m/sec),
 the experimental verification of above acoustic
 Hawking temperature will be rather difficult. However, the higher
 dimensional black hole analogue metric presents us an
 alternative way to detect extra dimensions other than in LHC. By detecting signals via
 the acoustic black hole's evaporation to brane-localized modes,
 one can in principle determine the exact dimensions of space-time.
 However, since the Standard Model particles must to be located to
 an ordinary 4-dimensional spacetime, extra dimensions can only be
 probed through the gravitational force. The continuity equation
 ${\nabla}\cdot \vec{v}=0$ in higher dimensions should only
 apply to gravitational waves or some unknown scalar particles.

\section{Bulk scalar emission: S wave and $l\geq 0$ }
\hspace*{7.5mm}In this section and the next section, we shall
calculate the decay rate of higher dimensional acoustic black holes
from the near-horizon low energy dynamics, which is expected to give
an experimental suggestion of detecting
the thermal radiation.\\
\hspace*{7.5mm} The radiation of higher dimensional acoustic black
holes is usually described as thermal spectrum in character with a
temperature $T_{4+n}$. The energy emitted per unit time (power
spectrum) by gravity-wave black hole analogues for a higher number
of dimensions can be given by
\begin{equation}
\frac{d E(\omega)}{dt}=\sum_{l}\sigma_{l,n}(\omega)\frac{\omega}{exp
(\omega/T_{4+n})- 1}\frac{d^{n+3}k}{(2\pi)^{n+3}},
\end{equation}where $l$ is the angular momentum quantum number and
$\mid k\mid=\omega$ for massless particles. However, considering the
nontrivial metric in the region exterior to the horizon, there
exists an effective potential barrier in this exterior region. This
potential barrier backscatters a part of the outgoing radiation back
into the black hole. Thus the original blackbody radiation is
modified by a frequency dependent filtering function
$\sigma_{l}({\omega})$, caused by the gravitational potential of the
black hole, which is called the ``greybody factor". Greybody factors
are important theoretically and experimentally in that they depend
on the number of extra dimensions and encode information on the
near-horizon structure of black holes and can be used to identify a
black hole event.\\
\hspace*{7.5mm}For S wave bulk scalar emission of the
$(4+n)$-dimensional gravity-wave black hole analogues, the
calculation is similar to that of higher dimensional Schwarzschild
cases. The calculations do not depend on the concrete form of the
metric. In fact, it has been proved that all spherically symmetric
black holes, regardless of the theory in which they arise, the low
energy cross section for massless minimally coupled scalars is
always the area of the horizon~\cite{das}. Following the method of
Ref.~\cite{kanti}, we can obtain the spherical wave absorption
probability,
\begin{equation}
\mid \mathcal{A}(\omega) \mid^{2}=\frac{2(\omega
r_{0})^{n+2}sin[\pi(n+1)/2]}{2^{n}(n+1)}\frac{\Gamma(\frac{1-n}{2})}{\Gamma(\frac{3+n}{2})}\end{equation}
The greybody factor can be computed by first evaluating the
absorption probability, $\mid \mathcal{A}(\omega) \mid^{2}$, from
the ratio of the in-going flux at the future horizon to the incoming
flux from past infinity with boundary condition that there is no
outgoing flux at the horizon, and then using the generalized
($4+n$)-dimensional optical theorem relation~\cite{gubser}
\begin{equation}
\sigma_{l}(\omega)=\frac{2^{n}\pi^{(n+1)/2}\Gamma(\frac{n+1}{2})}{n!\omega^{n+2}}\frac{(2l+n+1)(l+n)!}{l!}\mid
\mathcal{A}(\omega) \mid^{2}
\end{equation}
between the absorption cross section $\sigma_{l}({\omega})$ and the
absorption probability $\mid \mathcal{A}(\omega) \mid^{2}$ for the
\textit{l-}th partial waves.\\
\hspace*{7.5mm}We will then derive the scalar decay modes which are
not spherically symmetric, $l\neq 0$. We start with the
$4+n$-dimensional black hole analogue by rewriting
Eq.(\ref{acoustic}) in the following form,
\begin{equation}ds^{2}=-h(r)dt^{2}+h(r)^{-1}dr^{2}+r^{2}d\Omega_{n+2}^{2},\end{equation}
where $h(r)=1-\left(\frac{r_{0}}{r}\right)^{2n+4}$ and we have
assumed $\rho_{0}=c=1$. The scalar wave equation in this background
is separable if we make the ansatz
$\phi(t,r,\theta,\varphi)=e^{-i\omega t}R_{wl}(r)Y_{l}(\Omega)$,
where $Y_{l}(\Omega)$ are now the $(3+n)$-spatial dimensional
spherical harmonic functions. We can then obtain the radial part
equation by substituting the above ansatz into the Eq.(\ref{klein}),
which reads,
\begin{equation}
\label{radial}
\frac{h(r)}{r^{n+2}}\frac{d}{dr}\left[h(r)r^{n+2}\frac{dR}{dr}\right]+\left[\omega^{2}-\frac{h(r)l(l+n+1)}{r^{2}}\right]R=0.
\end{equation}
The idea of ~\cite{unr} is to solve this equation approximately in
three regions: near-horizon regions, far field regions and
intermediate regions, and match the solution across the boundaries
of the regions. We will use the same method here, but simply solve
the radial equation in near-horizon and far field regions, and
assume that keep only the lowest order terms in $\omega$ in each
region is enough. From the change of variables $0\leq r\rightarrow
h\leq 1$, we can write the scalar field equation Eq.(\ref{radial})
in the form,
\begin{eqnarray}
&&
h(1-h)\frac{d^{2}R}{dh^{2}}+\left(1-\frac{3n+7}{2n+4}h\right)\frac{dR}{dh}
\nonumber\\&& +\left[\frac{\omega^{2} r_{0}^{2}}{(2n+4)^{2}h
(1-h)}-\frac{l(l+n+1)}{(2n+4)^{2} (1-h)}\right]R=0,
\end{eqnarray} where we set $(\omega r)^{2}$ to be $(\omega
r_{0})^{2}$ near the horizon. By redefining
$R(h)=h^{\alpha}(1-h)^{\beta}F(h)$ and removing singularities at
$h=0$ and $h=1$, the above equation can be reduced to a
hypergeometric equation with $a=\alpha+\beta+\frac{n+3}{2n+4}$,
$b=\alpha+\beta$ and $c=2\alpha+1$, where,
\begin{eqnarray}\alpha_{\pm}=\pm\frac{i\omega
r_{0}}{2n+4},~~~~~~ \beta=\frac{1}{2}\pm
\frac{1}{2(n+2)}\sqrt{(l+n+2)^2-4\omega^{2} r_{0}^{2}}.
\end{eqnarray}
Thus, we have,
\begin{eqnarray}
\label{hyper}
h(1-h)\frac{d^{2}F}{dh^{2}}+\left[c-(1+a+b)h\right]\frac{dF}{dh}-abF=0,
\end{eqnarray}
which has a solution the hypergeometric function
$F(a,b,c;h)$~\cite{mab}. The criterion for the convergence of the
hypergeometric function demands that $Re(c-a-b)>0$, which force us
to choose $\beta=\beta_{-}$. Then, the general solution of
Eq.(\ref{ref}) is,
\begin{eqnarray}
\label{rh}R_{NH}(h)&=&A_{-}h^{\alpha_{\pm}}(1-h)^{\beta}F(a,b,c ;h)
\nonumber\\&+&
A_{+}h^{\pm\alpha_{\pm}}(1-h)^{\beta}F(a-c+1,b-c+1,2-c
;h)
\end{eqnarray}Expanding the above solution in the near-horizon region in the limit $r\rightarrow
r_{0}$, or $h\rightarrow 0$, and choosing $\alpha=\alpha_{-}$, we
obtain the result,
\begin{eqnarray}
\label{nearhorizon}
R_{NH}(h)=(\frac{r_{0}}{r})^{2\beta(n+2)}\left[A_{-}e^{(-i\omega
r_{0}^{n+2}y)}+A_{+}e^{(i\omega r_{0}^{n+2}y)}\right],
\end{eqnarray} where $y$ is defined by $y=\frac{\ln h(r)}{r^{n+1}_{0}(n+1)}$. To calculate the greybody factor, we must impose the
boundary condition that near the horizon the solution is purely
ingoing and then we set $A_{+}=0$~\cite{kmr}.\\
\hspace*{7.5mm}By using the $h\rightarrow (1-h)$ transformation of
hypergeometric functions,
\begin{eqnarray}
\label{hyperrelation}
&&F(a,b,c;h)=\frac{\Gamma(n)\Gamma(c-a-b)}{\Gamma(c-a)\Gamma(c-b)}F(a,b;a+b-c+1;1-h)\nonumber\\&&
+(1-h)^{c-a-b}\frac{\Gamma(c)\Gamma(a+b-c)}{\Gamma(a)\Gamma(b)}F(c-a;c-b,c-a-b+1;1-h),
\end{eqnarray}
the near-horizon solution (\ref{rh}) expanded in terms of $1-h$ is
given by,
\begin{eqnarray}
&&\label{nearh}R_{NH}(h)=A_{-}h^{\alpha}\left[(1-h)^{\beta}\frac{\Gamma(1+2\alpha)\Gamma(1-2\beta-\frac{n+3}{2n+4})}
{\Gamma(1+\alpha-\beta-\frac{n+3}{2n+4})\Gamma(1+\alpha-\beta)}F(a,b,a+b-c+1
; 1-h)~~~~~~~\right.\nonumber\\&& \left.
+(1-h)^{1-\beta-\frac{n+3}{2n+4}}\frac{\Gamma(1+2\alpha)\Gamma(2\beta+\frac{n+3}{2n+4}-1)}
{\Gamma(\alpha+\beta+\frac{n+3}{2n+4})\Gamma(\alpha+\beta)}F(c-a,c-b,c-a-b+1;
1-h)\right],
\end{eqnarray}where $A_{+}$ has been set to vanish. Expanding the above expression in the limit $h\rightarrow
1$, we get,
\begin{eqnarray}
\label{h1}R_{NH}(h)=&&A_{-}(\frac{r}{r_{0}})^{l}\frac{\Gamma(1+2\alpha)\Gamma(1-2\beta-\frac{n+3}{2n+4})}
{\Gamma(1+\alpha-\beta-\frac{n+3}{2n+4})\Gamma(1+\alpha-\beta)}
~~~~~~~\nonumber\\&&
+(\frac{r_{0}}{r})^{l+n+1}\frac{\Gamma(1+2\alpha)\Gamma(2\beta+\frac{n+3}{2n+4}-1)}
{\Gamma(\alpha+\beta+\frac{n+3}{2n+4})\Gamma(\alpha+\beta)}.
\end{eqnarray}
\hspace*{7.5mm}The derivation of the far-field zone which is defined
by $r\gg r_{0}$. In this limit, $h(r)\simeq 1$ and, by setting
$R(r)=f(r)/r^{(n+1)/2}$, Eq.(\ref{radial}) can be rewritten as
\begin{eqnarray}
\frac{d^{2}f}{dr^{2}}+\frac{1}{r}\frac{df}{dr}+\left[\omega^{2}-\frac{(2l+n+1)^{2}}{4r^{2}}\right]f=0,
\end{eqnarray}
which is a $(l+\frac{n+1}{2})$-th order Bessel equation. The
solution of the above equation are the Bessel functions
\begin{eqnarray}
\label{ff}R_{FF}=\frac{B_{+}}{r^{(n+1)/2}}J_{l+\frac{n+1}{2}}(\omega
r)+\frac{B_{-}}{r^{(n+1)/2}}Y_{l+\frac{n+1}{2}}(\omega r),
\end{eqnarray}where $J_{l+\frac{n+1}{2}}(\omega
r)$ and $Y_{l+\frac{n+1}{2}}(\omega r)$ are the Bessel functions of
the first and second kind, respectively.  For small $\omega r\ll 1$,
we can expand the above formula into,
\begin{eqnarray}
\label{far}&&R_{FF}\simeq
\frac{B_{+}r^{l}}{\Gamma(l+\frac{n+3}{2})}\left(\frac{\omega}{2}\right)^{l+\frac{n+1}{2}}
-\frac{B_{-}}{r^{l+n+1}}
\left(\frac{2}{\omega}\right)^{l+\frac{n+1}{2}}\frac{\Gamma(l+\frac{n+1}{2})}{\pi}
\end{eqnarray}
In order to match the solutions across the boundaries of
near-horizon field and far-field zone, we need to rewrite the
near-horizon solution in terms of ($1-h$), before expanding the
solution  in the limit $r\gg r_{0}$. Matching the two solutions
Eqs.(\ref{h1}) and (\ref{far}), we obtain the ratio,
\begin{eqnarray}
&&\frac{B_{+}}{B_{-}}\approx - \left(\frac{2}{\omega
r_{0}}\right)^{2l+n+1} \nonumber\\&&
\frac{\Gamma(l+\frac{n+1}{2})^{2}(l+\frac{n+1}{2})
\Gamma(1-2\beta-\frac{n+3}{2n+4})\Gamma(\alpha+\beta)\Gamma(\alpha+\beta+\frac{n+3}{2n+4})}{\pi
\Gamma(1+\alpha-\beta)\Gamma(1+\alpha-\beta-\frac{n+3}{2n+4})\Gamma(2\beta+\frac{n+3}{2n+4}-1)}.
\end{eqnarray}
\begin{figure}
\psfig{file=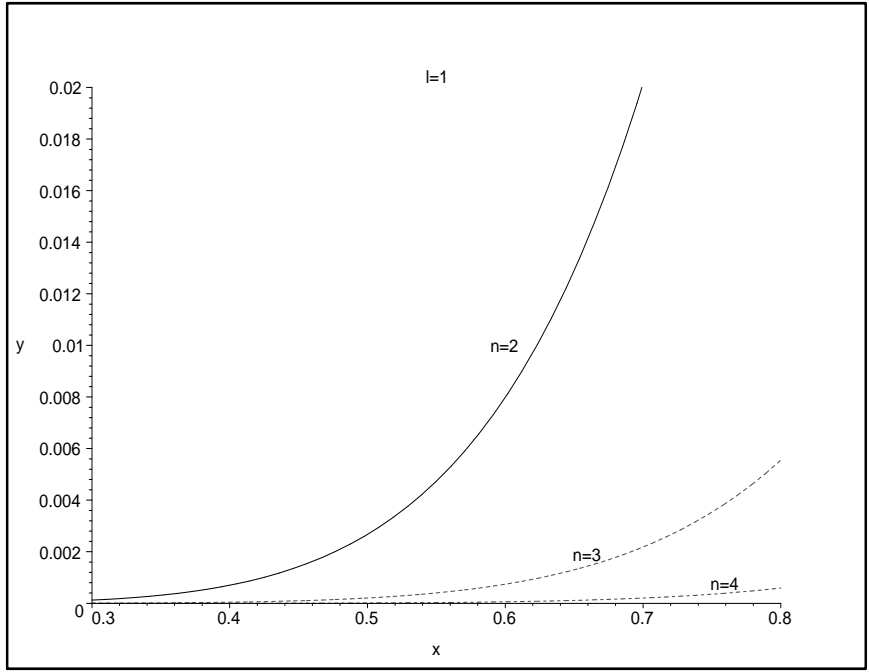 ,height=4.5in,width=3.5in,angle=270 }\caption{
Analytical results for the absorption probability for a
$(4+n)$-dimensional bulk scalar field for $l=1$, where $x=\omega
r_{0}$ and $y=\mid \mathcal{A} \mid^{2}$. }
\end{figure}
\begin{figure}
\psfig{file=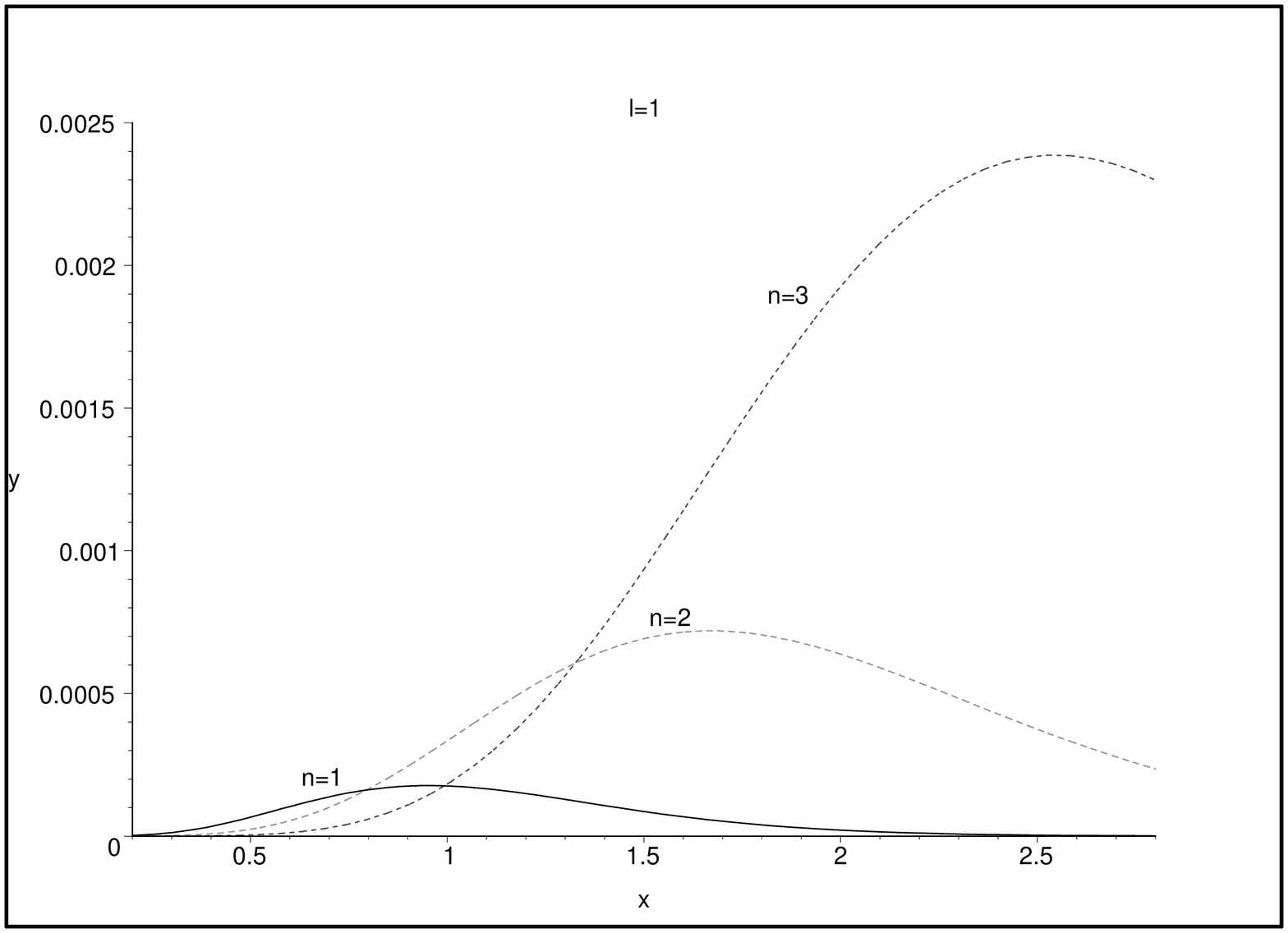 ,height=4.5in,width=3.5in,angle=270 }\caption{
Analytical results for the energy rates for scalars from a
$(4+n)$-dimensional black hole in the bulk for $l=1$,  where
$x=\omega r_{0}$ and $y=\frac{d^{2}E}{dt d\omega}[r_{0}^{-1}]$. }
\end{figure}
The ingoing and outgoing waves of Eq.(\ref{far}) can be decomposed
by introducing the redefinition of amplitudes. We expand
Eq.(\ref{far}) in the limit $r\rightarrow \infty$, and then we
obtain,
\begin{eqnarray}
R^{(\infty)}=A^{(\infty)}_{in}\frac{e^{-i\omega
r}}{\sqrt{r^(n+2)}}+A^{(\infty)}_{out}\frac{e^{i\omega
r}}{\sqrt{r^(n+2)}},
\end{eqnarray}
where $A^{(\infty)}_{in}$ and $A^{(\infty)}_{out}$ is defined by,
\begin{eqnarray}
A^{(\infty)}_{in}&=&\frac{B_{+}+iB_{-}}{\sqrt{2\pi\omega}}e^{i\pi(l+\frac{n}{2}+1)/2},\nonumber\\
A^{(\infty)}_{out}&=&\frac{B_{+}-iB_{-}}{\sqrt{2\pi\omega}}e^{-i\pi(l+\frac{n}{2}+1)/2}
\end{eqnarray}
The reflection coefficient $\mathcal{R}$ is defined as the ratio of
the outgoing amplitude over the incoming amplitude at infinity,
\begin{eqnarray}
\mathcal{R}=\left|\frac{A^{(\infty)}_{out}}{A^{(\infty)}_{in}}\right|^2.
\end{eqnarray}
The absorption probability can be written, in terms of
$B=B_{+}/B_{-}$, as
\begin{equation}
\mid \mathcal{A}(\omega)
\mid^{2}=1-\mid\mathcal{R}\mid^{2}=\frac{2i(B^{*}-B)}{BB^{*}+i(B^{*}-B)+1}.
\end{equation} In the limit of $wr_{0}\ll 1$ and $BB^{*}\gg i(B^{*}-B)\gg
1$, the absorption probability can approximately  be written as
\begin{eqnarray}
\mid \mathcal{A}(\omega) \mid^{2}\simeq &&
\frac{8\pi(2n+5-2l)}{(l+\frac{n+1}{2})(2l-1)^{2}}\times
\nonumber\\&& (\frac{\omega r_{0}}{2})^{2l+n+2}
\frac{\Gamma(\frac{3}{4}+\frac{l}{2n+4})^{2}\Gamma(\frac{n}{4(n+2)}+\frac{l}{2n+4})^{2}}
{\Gamma(l+\frac{n+1}{2})^{2}\Gamma(\frac{2l-1}{2n+4})^{2}}
\end{eqnarray}
\hspace*{7.5mm}From Figure 1, we can see that if we fix the angular
momentum number and vary only the number of extra dimensions, the
absorption probability decreases as $n$ increases, since the
expansions of $\mid \mathcal{A} \mid^{2}$ is in powers of $\omega
r_{0}\ll 1$. Thus, $\mid \mathcal{A} \mid^{2}$ should become more
and more suppressed as $n$ increases. The same behavior is observed
if we fix instead $n$ and vary $l$.\\
\hspace*{7.5mm}According to Figure 2, the emission rate of scalar
fields in the bulk is enhanced as the number of extra dimensions
increases. This is caused by the increase of the temperature of
gravity-wave black hole analogues, which finally overcome the
decreases in the value of the greybody factor and causes the
enhancement of the emission rate with $n$ at high energies.
\section{Brane-localized scalar emission for $l\geq 0$}
\hspace*{7.5mm}If the acoustic black hole is formed from matter on
the brane, the scalar field is confined on the 3-brane embedded in a
$(4+n)$-dimensional space-time. The induced metric on the brane will
be,
\begin{equation}ds^{2}=-h(r)dt^2+h(r)^{-1}dr^{2}+r^{2}(d\theta^{2}+\sin^{2}\theta d\varphi^2),\end{equation}
where $h(r)=1-(\frac{r_{0}}{r})^{2(n+2)}$. on the brane, the event
horizon is still at $r=r_{0}$ and its area is $A_{4}=4\pi r_{0}^2$.
This induced acoustic metric on the brane is certainly not the
$4$-dimensional acoustic geometry. The calculation of Hawking
radiation relies on mainly on properties of the horizon, such as its
surface gravity. We shall calculate the absorption cross-section of
the brane-localized scalar field, which according to
Ref.\cite{emparan}, most of the energy radiated by black holes goes
into modes on the brane. Using the separation of variables,
$\phi(t,r,\theta,\varphi)=e^{-i\omega t}R_{wl}(r)Y_{l}(\Omega)$,
where $Y_{l}(\Omega)$ are now the usual three-dimensional spherical
harmonic functions, the radial equation of Eq.(\ref{klein}) is
written as,\begin{equation} \label{braneeq}
\frac{h(r)}{r^{2}}\frac{d}{dr}\left[h(r)r^{2}\frac{dR}{dr}\right]
+\left[\omega^{2}-\frac{h(r)}{r^{2}}l(l+1)\right]R=0\end{equation}
Similar to the discussions in section 3, we solve this equation in
two regions: the near-horizon region and far-field region.
 Our starting point is the Klein-Gordon equation under the
 near-horizon metric background.
 In terms of $h$, the radial differential equation now takes the
form,
\begin{eqnarray}
&&h(1-h)\frac{d^{2}R}{dh^2}+\left[1-\frac{4n+7}{2(n+2)}\right]\frac{dR}{dh}
~~~~\nonumber\\
&&+\left[\frac{\omega^{2}r^{2}}{(2n+4)^{2}h(1-h)}-\frac{l(l+1)}{(2n+4)^{2}(1-h)}\right]R=0
\end{eqnarray}
If we further define $R(h)=h^{\alpha}(1-h)^{\beta}F(h)$ to remove
singularities at points $h=0$ and $h=1$, the above equation assumes
the standard form of a hypergeometric equation
\begin{eqnarray}
\label{hypergeo}
h(1-h)\frac{d^{2}F}{dh^{2}}+\left[c-(1+a+b)h\right]\frac{dF}{dh}-abF=0,
\end{eqnarray}
 with indices
$a=\alpha+\beta+\frac{2n+3}{2n+4}$,$b=\alpha+\beta$ and
$c=1+2\alpha$, where $\alpha_{\pm}=\frac{i\omega r}{2n+2}$ and
$\beta=\frac{1}{4(n+2)}(1\pm \sqrt{(2l+1)^{2}-4\omega^{2} r^{2}})$.
The criterion for the convergence of the horizon and imposing the
boundary condition that only incoming waves exist near $r\simeq
r_{0}$, one can find that $A_{+}=0$ for $\alpha=\alpha_{-}$. To
express the form of the solution for small $h$, we express the
$F(h)$ in terms of $1-h$, by using the hypergeometric relation
Eq.(\ref{hyperrelation}). Thus after expanding $R_{NH}(h)$ for $r\gg
r_{0}$, we find that the desired solution for near-horizon region is
\begin{eqnarray}
\label{noh}
 &&R_{NH}(h)\simeq
A_{-}(\frac{r}{r_{0}})^{l}\frac{\Gamma(1+2\alpha)
\Gamma(1-2\beta-\frac{2n+3}{2n+4})}{\Gamma(1+\alpha-\beta-\frac{2n+3}{2n+4})\Gamma(1+\alpha+\beta)}
~~~~~\nonumber\\
&&+A_{-}(\frac{r_{0}}{r})^{l+1}\frac{\Gamma(1+2\alpha)\Gamma(2\beta+\frac{2n+3}{2n+4}-1)}
{\Gamma(\alpha-\beta+\frac{2n+3}{2n+4})\Gamma(\alpha+\beta)}
\end{eqnarray}
For the far-field region which is defined by $r\gg r_{0}$,  noting
that $h(r)\rightarrow 1$ and setting $R(r)=f(r)/r^{1/2}$, we rewrite
Eq.(\ref{braneeq}) as,
\begin{eqnarray}
\frac{d^2f}{dr^2}+\frac{1}{r}\frac{df}{dr}+\left[\omega^2-\frac{(2l+1)^2}{4r^2}\right]f=0.
\end{eqnarray}
The far field region solution can be expressed in terms of the
Bessel function $J_{l+1/2}(\omega r)$ and Neumann function
$Y_{l+1/2}(\omega r)$,
\begin{eqnarray}
\label{farfield}R_{FF}=\frac{B_{+}}{r^{1/2}}J_{l+\frac{1}{2}}(\omega
r)+\frac{B_{-}}{r^{1/2}}Y_{l+\frac{1}{2}}(\omega r),
\end{eqnarray}
 At this stage, we expand the general solution Eq.(\ref{farfield}),
in the low energy limit $\omega r\ll 1$ and find that,
\begin{eqnarray}
\label{final} &&R_{FF}\simeq
\frac{B_{+}r^{l}}{\Gamma(l+3/2)}\left(\frac{\omega}{2}\right)^{l+1/2}
 -\frac{B_{-}}{r^{l+1}}
\left(\frac{2}{\omega}\right)^{l+1/2}\frac{\Gamma(l+1/2)}{\pi}
\end{eqnarray}Matching the solution Eq.(\ref{noh}) with (\ref{final}), we obtain the
ratio,
\begin{eqnarray}
&&\frac{B_{+}}{B_{-}}=- \left(\frac{2}{\omega r_{0}}\right)^{2l+1}
\nonumber\\&&
 \frac{\Gamma(l+\frac{1}{2})^{2}(l+\frac{1}{2})
\Gamma(1-2\beta-\frac{2n+3}{2n+4})\Gamma(\alpha+\beta)\Gamma(\alpha+\beta+\frac{2n+3}{2n+4})}{\pi
\Gamma(1+\alpha-\beta)\Gamma(1+\alpha-\beta-\frac{2n+3}{2n+4})\Gamma(2\beta+\frac{2n+3}{2n+4}-1)}
\end{eqnarray}
We can now compute the absorption probability $\mid
\mathcal{A}(\omega) \mid^{~2}$ in the low energy limit $\omega
r_{0}\ll 1$, which goes as,
\begin{eqnarray}
\mid \mathcal{A}(\omega)
\mid^{2}=\frac{16\pi}{(2l+1)^{2}}\left(\frac{\omega
r_{0}}{2}\right)^{2l+2}\frac{\Gamma(\frac{l+1}{2n+4})^{2}\Gamma(1+\frac{l}{2n+4})^{2}}
{\Gamma(l+\frac{1}{2})^{2}\Gamma(1+\frac{2l+1}{2n+4})^{2}}
\end{eqnarray}
\hspace*{7.5mm}Figure 3 demonstrates that if we keep $n$ fixed and
varying $l$, the absorption probability decreases as $l$ increases.
The dominant term becomes more and more suppressed by extra powers
of $wr_{0}$ and its numerical coefficients also decreases. If we fix
instead $l$ and vary $n$, a different behavior from the one observed
in the case with a bulk scalar field emerges, that is to say the
leading term remains the same since it is $n$-independent. Figure 4
is to compare the results of the absorption probability derived in
the case of a brane-localized $(n>0)$ scalar field with a purely
4-dimensional $(n=0)$ scalar field. For higher partial waves
$(l>0)$, the value of the absorption probability in the case of a
brane-localized $(n>0)$ scalar field is larger than the one for a
purely 4-dimensional $(n=0)$ field, while for $l=0$ they are the
same.
\begin{figure}
\psfig{file=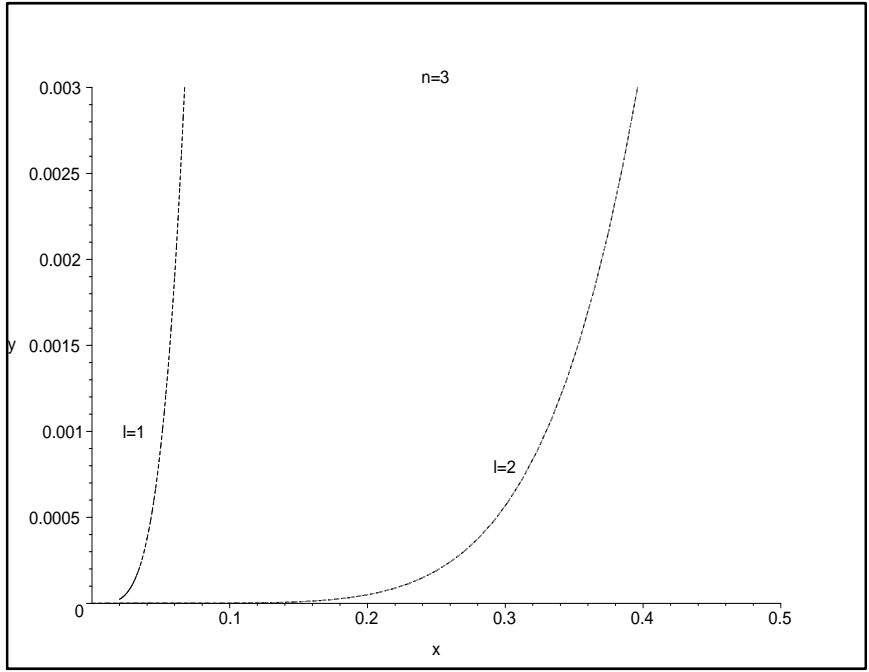 ,height=4.5in,width=3.5in,angle=270 }\caption{
Analytical results for the absorption probability for a $(4+n)$
black holes analogues on the brane for $n=3$, where $x=\omega r_{0}$
and $y=\mid \mathcal{A} \mid^{2}$. }
\end{figure}
\begin{figure}
\psfig{file=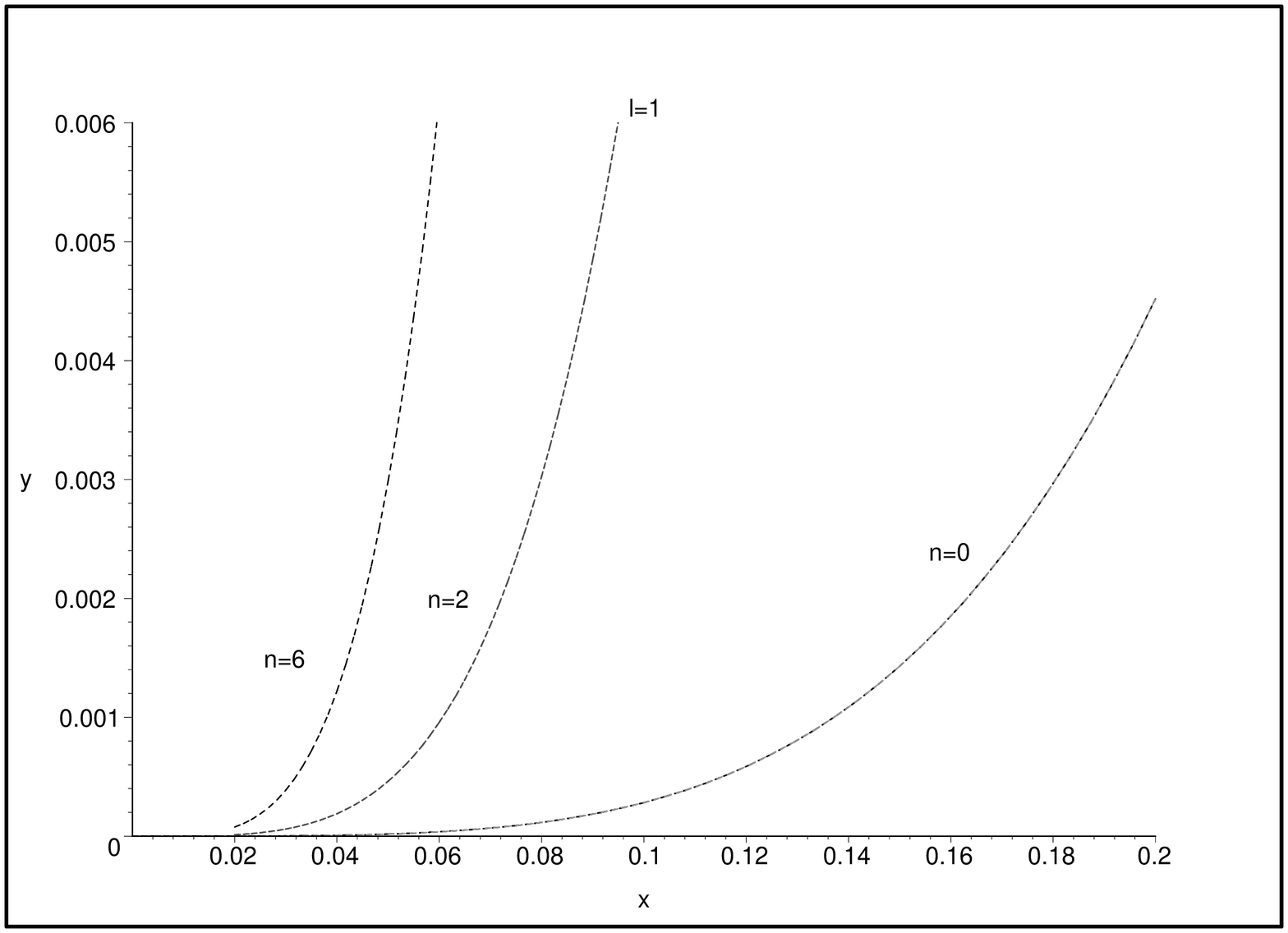 ,height=4.5in,width=3.5in,angle=270
}\caption{ Analytical results for the absorption probability for a
$(4+n)$ black holes analogues on the brane for $l=3$, where
$x=\omega r_{0}$ and $y=\mid \mathcal{A} \mid^{2}$. }
\end{figure}
\begin{figure}
\psfig{file=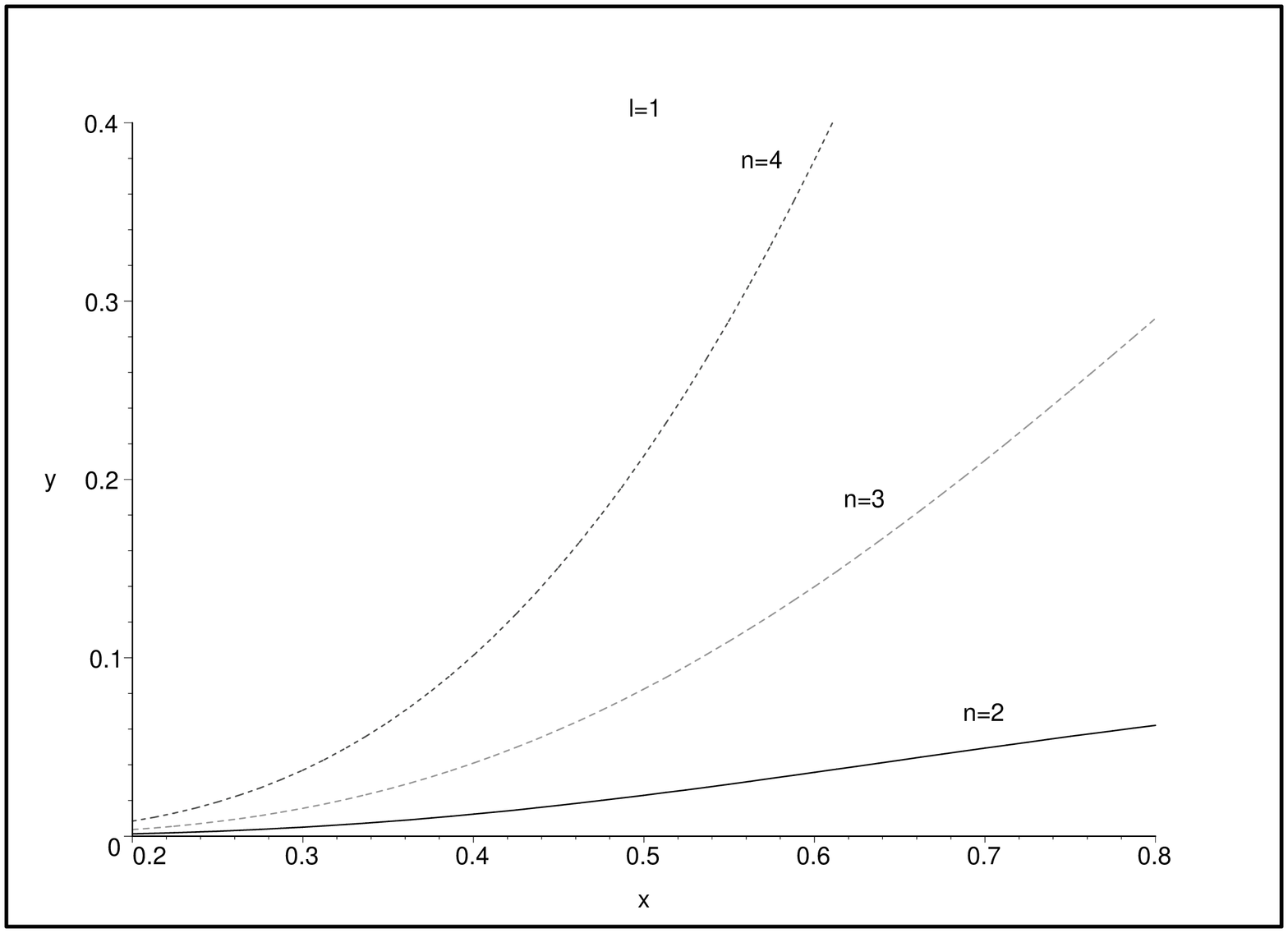 ,height=4.5in,width=3.5in,angle=270
}\caption{ Analytical results for the energy rates for scalars from
a $(4+n)$-dimensional black hole on the brane for $l=1$, where
$x=\omega r_{0}$ and $y=\frac{d^{2}E}{dt d\omega}[r_{0}^{-1}]$. }
\end{figure}
 Figure 5 depicts the behavior
of the energy emission rates for particles with the angular number
$l=1$ in the low and intermediate energy regime. The figure shows
that the energy and the number of particles, emitted per unit time
and energy interval is strongly enhanced, as $n$ increases since the
temperature of the black hole is given by the relation
$T_{4+n}\propto(n+2)/2\pi r_{0}$, which indicates that for fixed
$r_{0}$, the temperature of the gravity-wave black hole analogues
increases as $n$ increases. This means the energy of the emission
particles also increases.
\section{Conclusions}
\hspace*{7.5mm}In summary, we have  extended the 4-dimensional
acoustic black hole metric to higher dimensions that is similar to
higher dimensional Schwarzschild metric in form but is not exactly
any of the standard geometries typically considered in general
relativity. The fluids here have been assumed to fill all the
spacial dimensions including extra dimensions. We emphasize that the
higher dimensional acoustic black holes discussed above is just a
model, which may help us understand the physics beyond the standard
model and are falsifiable by the future experiments. Although it
might be difficult to detect the Hawking effects of these analogues
in experiments since the Hawking temperature in fact is very low,
they provide us an otherwise way to probe extra
dimensions.\\
\hspace*{7.5mm}The scalar emission of Hawking particles in both
$(4+n)$-dimensional bulk scalar field and 4-dimensional
brane-localized scalar field were studied respectively in a higher
dimensional gravity-wave black hole analogues background. The
amplitude probability in a bulk scalar field was obtained that
allows one to find its dependence on the number of extra dimensions
$n$ and the angular momentum number $l$. We found that if we fix the
angular momentum number and vary only the number of extra
dimensions, the absorption probability decreases as $n$ increases,
and $\mid \mathcal{A} \mid^{2}$ should become more and more
suppressed as $n$ increases. The same behavior is observed if we fix
instead $n$ and vary $l$. The case in which the scalar field is
confined on a 3-brane in a higher dimensional spacetime background
is also discussed. The resulting absorption probability depends only
on the angular momentum number through $(\omega r_{0})^{2l+2}$. If
we keep $n$ fixed and varying $l$, the absorption probability
decreases as $l$ increases. But, in both bulk and brane-localized
cases, the energy emission rates are enhanced as $n$ increases since
the temperature of the gravity-wave black hole
analogues increases as $n$ increases.\\
{\textbf{Acknowledgments}}\\~~~~~  \hspace*{7.5mm}S. W. Kim is
supported in part by KRF.

\end{document}